\documentclass[twocolumn,aps,prl,showpacs,amsmath,floatfix,
amssymb,superscriptaddress]{revtex4}

\usepackage{graphicx}
\usepackage{dcolumn}
\usepackage{bm}

\begin{document}

\title{Magnetic-field induced superconductor-antiferromagnet transition
in lightly doped $R$Ba$_2$Cu$_3$O$_{6+x}$ ($R$$\,=\,$Lu, Y)
crystals}

\author{A. N. Lavrov}
\affiliation{Institute of Inorganic Chemistry, Lavrentyeva-3,
Novosibirsk 630090, Russia}
\author{L. P. Kozeeva}
\affiliation{Institute of Inorganic Chemistry, Lavrentyeva-3,
Novosibirsk 630090, Russia}
\author{M. R. Trunin}
\affiliation{Institute of Solid State Physics, Institutskaya-2,
Chernogolovka 142432, Moscow region, Russia}
\author{V. N. Zverev}
\affiliation{Institute of Solid State Physics, Institutskaya-2,
Chernogolovka 142432, Moscow region, Russia}

\date{\today}

\begin{abstract}

The remarkable sensitivity of the $c$-axis resistivity and
magnetoresistance in cuprates to the spin ordering is used to clarify
the doping-induced transformation from an antiferromagnetic (AF)
insulator to a superconducting (SC) metal in $R$Ba$_2$Cu$_3$O$_{6+x}$
($R$$\,=\,$Lu, Y) single crystals. The established phase diagram
demonstrates that the AF and SC regions apparently overlap: the
superconductivity in $R$Ba$_2$Cu$_3$O$_{6+x}$, in contrast to
La$_{2-x}$Sr$_x$CuO$_4$, sets in before the long-range AF order is
completely destroyed by hole doping. Magnetoresistance measurements of
superconducting crystals with low $T_c \,$$\leq \,$$15-20\,$K give a
clear view of the magnetic-field induced superconductivity suppression
and recovery of the long-range AF state. What still remains to be
understood is whether the AF order actually persists in the SC state or
just revives when the superconductivity is suppressed, and, in the
former case, whether the antiferromagnetism and superconductivity reside
in nanoscopically separated phases or coexist on an atomic scale.

\end{abstract}

\maketitle


\section{Introduction}

The transformation, upon charge doping, of an antiferromagnetic (AF)
Mott insulator into a superconducting (SC) metal and the role of AF
correlations in the appearance of superconductivity have challenged
researchers since the discovery of high-$T_c$ superconductivity in
cuprates. Is the AF order an indispensable component or a competitor for
the high-$T_c$ phenomenon? In a prototype high-$T_c$ cuprate
La$_{2-x}$Sr$_x$CuO$_4$, the long-range AF order is destroyed by doped
holes way before the superconductivity sets in \cite{RMP_LSCO}, which
has led to a general belief that the spin frustration is a prerequisite
for metallic conduction and superconductivity. The destructive impact of
static spin order on superconductivity was further supported by the
observation of SC suppression at a peculiar 1/8 doping in
La$_{2-x}$Ba$_x$CuO$_4$ \cite{1_8}. On the other hand, spin excitations
are often suggested to provide glue for SC pairing, implying the
ultimate importance of AF correlations, be they static or dynamic.
Besides, the incompatibility of static AF order and SC may be not
necessarily a general feature of cuprates. In $R$Ba$_2$Cu$_3$O$_{6+x}$
($R$ is a rare-earth element), for instance, the long-range AF order
survives up to much higher doping levels than in La$_{2-x}$Sr$_x$CuO$_4$
\cite{YBCO_AF, diagram2, AF_SC, anis, diagram3}, though the possibility
of its coexistence with superconductivity still remains to be clarified.

In strongly anisotropic high-$T_c$ cuprates, the $c$-axis charge
transport appears to be remarkably sensitive to the spin ordering in
CuO$_2$ planes. In $R$Ba$_2$Cu$_3$O$_{6+x}$ crystals, for example, the
$c$-axis resistivity $\rho_c(T)$ exhibits a steep increase at the
N\'{e}el temperature $T_N$ \cite{AF_SC, anis, ourmag}. Even relatively
weak modifications of the spin structure such as spin-flop or
metamagnetic transitions result in surprisingly large changes---by up to
an order of magnitude---in the $c$-axis resistivity of both hole-doped
La$_{2-x}$Sr$_x$CuO$_4$ \cite{MR_WF, LSCO_MR} and electron-doped
Pr$_{1.3-x}$La$_{0.7}$Ce$_{x}$CuO$_4$ \cite{PLCCO_MR} and
Nd$_{2-x}$Ce$_{x}$CuO$_4$ crystals \cite{NCCO_MR}. This sensitivity of
the interplane charge transport in cuprates to the spin order can be,
and actually is, employed for tracing the evolution of the spin state
with doping, temperature, or magnetic fields \cite{MR_WF, AF_SC, anis,
ourmag, PLCCO_MR}.

While electrical resistivity measurements have proved to be a very
convenient tool for mapping the magnetic phase diagrams in cuprates,
their usage has an obvious limitation; namely, they fail as the
superconductivity sets in. Because of this limitation, previous
resistivity studies of $R$Ba$_2$Cu$_3$O$_{6+x}$ crystals \cite{AF_SC,
anis} could not clarify whether the long-range AF order vanishes by the
onset of superconductivity, or extends further, intervening the SC
region. It sounds tempting to employ strong magnetic fields to suppress
the superconductivity and to use the $c$-axis resistivity technique of
detecting the spin order in otherwise inaccessible regions of the phase
diagram. In the present paper, we use this approach to study the very
region of the AF-SC transformation in LuBa$_2$Cu$_3$O$_{6+x}$ and
YBa$_2$Cu$_3$O$_{6+x}$ single crystals.

\section{Experiment}

$R$Ba$_2$Cu$_3$O$_{6+x}$ single crystals with nonmagnetic rare-earth
elements $R$$\,=\,$Lu and Y were grown by the flux method and their
oxygen stoichiometry was tuned to the required level by high-temperature
annealing with subsequent quenching \cite{AF_SC, whisk}. In order to
ensure that no oxygen-enriched layer was formed at the crystal surface
during the quenching process, one of the crystals was dissolved in acid
in several steps; resistivity measurements detected no considerable
change in the SC transition upon the crystal's surface destruction. The
$c$-axis resistivity $\rho_c(T)$ was measured using the ac four-probe
technique. To provide a homogeneous current flow along the $c$-axis, two
current contacts were painted to almost completely cover the opposing
$ab$-faces of the crystal, while two voltage contacts were placed in
small windows reserved in the current ones \cite{AF_SC}. The
magnetoresistance (MR) was measured by sweeping temperature at fixed
magnetic fields up to 16.5 T applied along the $c$ axis of the crystals.

\section {Results and Discussion}

A representative $\rho_c(T)$ curve obtained for a
LuBa$_2$Cu$_3$O$_{6+x}$ single crystal with a doping level slightly
lower than required for the onset of superconductivity is shown in
Fig.~1. In general, the $c$-axis resistivity in $R$Ba$_2$Cu$_3$O$_{6+x}$
crystals of non-SC composition exhibits two peculiar features upon
cooling below room temperature, both of which can be seen in Fig.~1. The
first one is a pronounced crossover at $T_m$ ($T_m \,$$\approx
\,$$120\,$K for the particular composition in Fig.~1), indicating a
change---with decreasing temperature---of the dominating conductivity
mechanism from some kind of thermally activated hopping to a coherent
transport \cite{AF_SC, anis, ourmag, whisk, Trunin}. It is worth noting
that a similar coherent-incoherent crossover was observed in other
layered oxides as well \cite{Valla}. The second feature is a sharp
growth of the resistivity associated with the long-range AF ordering
\cite{AF_SC, anis, ourmag}. If the crystals were less homogeneous, the
low-temperature resistivity upturn would be easy to confuse with a usual
disorder-induced charge localization. However, this sharp resistivity
anomaly with a characteristic negative peak in the derivative (inset in
Fig.~1) is definitely related to the spin ordering at the N\'{e}el
temperature $T_N$: it has been traced from the parent compositions
$R$Ba$_2$Cu$_3$O$_{6}$ with well-known $T_N$ to avoid any doubt in its
origin.

\begin{figure}[!t]
\includegraphics*[width=7.6cm]{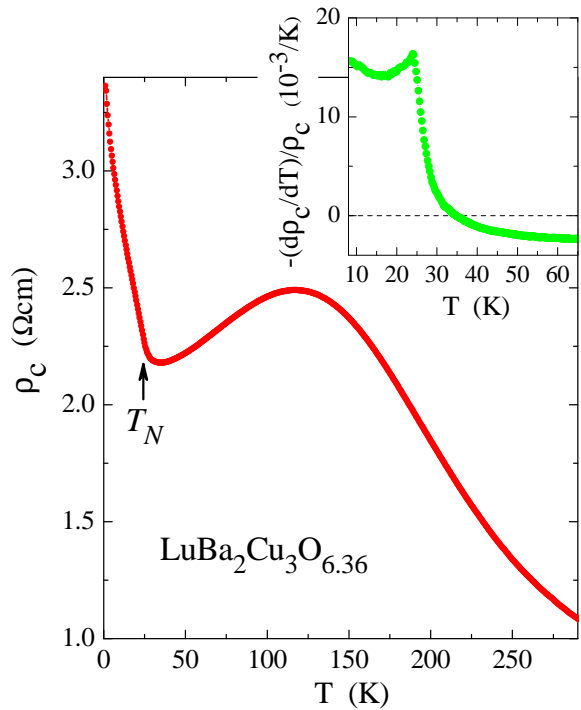}
\caption{Out-of-plane resistivity, $\rho_c(T)$, of a
LuBa$_2$Cu$_3$O$_{6.36}$ single crystal. The sharp growth of the
resistivity upon cooling below $\approx \,$$25 \,$K is caused by
the AF ordering. Inset: anomaly in the normalized derivative
$(d\rho_c/dT)/\rho_c$ associated with the N\'{e}el transition.}
\label{fig1}
\end{figure}

In carefully prepared crystals, the AF transitions remain sharp for all
compositions, including crystals with very low
$T_N\,$$\approx$$\,20\,$K, that is $\sim\,$20 times lower than original
$T_{N0}\,$$\approx \,$$420\,$K in parent $R$Ba$_2$Cu$_3$O$_{6}$. It is
important to emphasize that the transitions at $T_N\,$$\approx$$\,20\,$K
remain virtually as sharp as in undoped parent crystals even though spin
freezing into a spin-glass state is usually expected for such low
temperatures and high hole concentrations \cite{SG}. Moreover, the
impact of the AF ordering on $\rho_c(T)$ does not weaken with decreasing
N\'{e}el temperature; as can be seen in Fig.~1, the resistivity of the
crystal with $T_N\,$$\approx$$\,24\,$K increases by more than 50\% upon
cooling below $T_N$, while in crystals with $T_N\,$$>$$\,100\,$K the
corresponding $\rho_c$ growth does not exceed 15-20\% \cite{AF_SC,
ourmag}.

What do these observations tell about the impact of doped holes on
copper spins? Apparently, the sharp AF transitions are hard to reconcile
with strongly frustrated spin states and disordered spin textures in
CuO$_2$ planes that are usually expected to emerge in cuprates with
doping \cite{Gooding}. Besides, a strong frustration-induced reduction
of the staggered magnetization $M^{\dag}$ required to account for the
decrease in $T_N$ would necessarily diminish the impact of the
interplane spin ordering on $\rho_c$, which again disagrees with
observations. On the other hand, if the role of mobile doped holes is
not to introduce a uniform spin frustration, but simply to break the
long-range AF order into two-dimensional domains in CuO$_2$ planes, the
observed behavior is easier to understand. In this case, it is the AF
domains in CuO$_2$ planes that become the elementary magnetic units and
the long-range AF state should develop through ordering of their phases,
which can occur rather abruptly. Correspondingly, the $T_N$ evolution
with doping should be governed by the decreasing AF domain size, rather
than $M^{\dag}$. In turn, the ordering of AF domains whose local
staggered magnetization does nor change appreciably with hole doping
leaves room for large changes in the $c$-axis resistivity even at low
$T_N$.

The $T_N$ and $T_c$ values determined from the $c$-axis resistivity of
$R$Ba$_2$Cu$_3$O$_{6+x}$ crystals have been used to establish the
doping-temperature phase diagram in Fig.~2. A peculiarity of
$R$Ba$_2$Cu$_3$O$_{6+x}$ crystals is that their doping level is
determined both by the oxygen content and by the degree of its ordering
\cite{AF_SC, anis}. For characterizing the doping level, we use
therefore the in-plane conductivity $\sigma_{ab}(280 K)$ instead of the
oxygen content; the former is a good measure of the hole density given
that the hole mobility stays almost constant in the doping region under
discussion \cite{mobility}.

\begin{figure}[!t]
\includegraphics*[width=8.65cm]{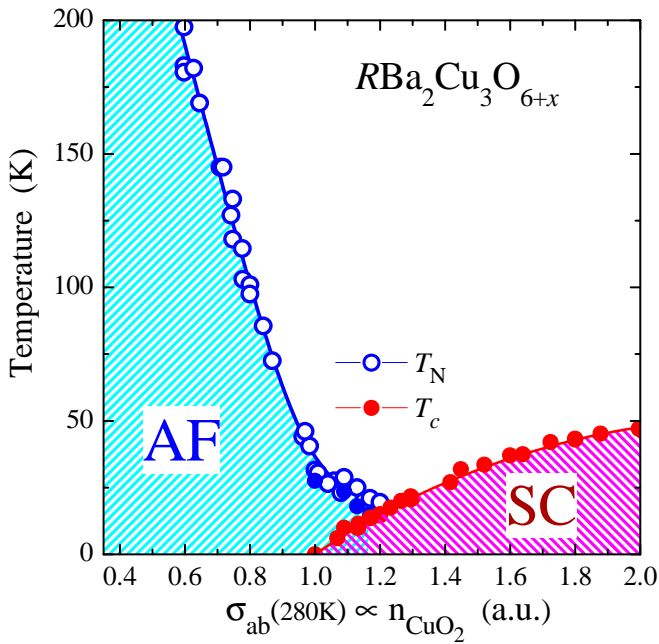}
\caption{Phase diagram of $R$Ba$_2$Cu$_3$O$_{6+x}$ ($R$ = Lu, Tm,
Y) crystals near the AF-SC transformation. The AF and SC transition
temperatures, $T_N$ and $T_c$, are presented as a function of the
in-plane conductivity $\sigma_{ab}(280 K)$ which is a good measure of
the hole density in the shown doping region \cite{mobility}. The
N\'{e}el temperature was determined either at the position of the jump
(middle point) in the derivative $d\rho_c/dT$ (open circles), or at the
position of the negative peak in $d\rho_c/dT$ (solid circles).}
\label{fig2}
\end{figure}

As can be seen in Fig.~2, the long-range AF order in
$R$Ba$_2$Cu$_3$O$_{6+x}$ appears to be much more stable than in
La$_{2-x}$Sr$_x$CuO$_4$ where it vanishes well in advance before the
onset of superconductivity. In $R$Ba$_2$Cu$_3$O$_{6+x}$, the AF phase
boundary flattens upon approaching the SC compositions and hits the SC
region, crossing the $T_c$ line at $\approx \,$$15\,$K. The observed
overlap of the SC and AF regions is very close to the area where $\mu$SR
studies of YBa$_2$Cu$_3$O$_{6+x}$ ceramics revealed the coexistence of
superconductivity with spontaneous static magnetism \cite{diagram2,
diagram3}. Given that the $c$-axis resistivity studied here is sensitive
to the {\it interlayer} spin ordering, the static magnetism detected by
$\mu$SR \cite{diagram2, diagram3} should in fact be related to the
three-dimensional AF order. The superconductivity in
$R$Ba$_2$Cu$_3$O$_{6+x}$ thus develops directly from the AF-ordered
state without any intervening paramagnetic or spin-glass region.

\begin{figure}[!t]
\includegraphics*[width=8.65cm]{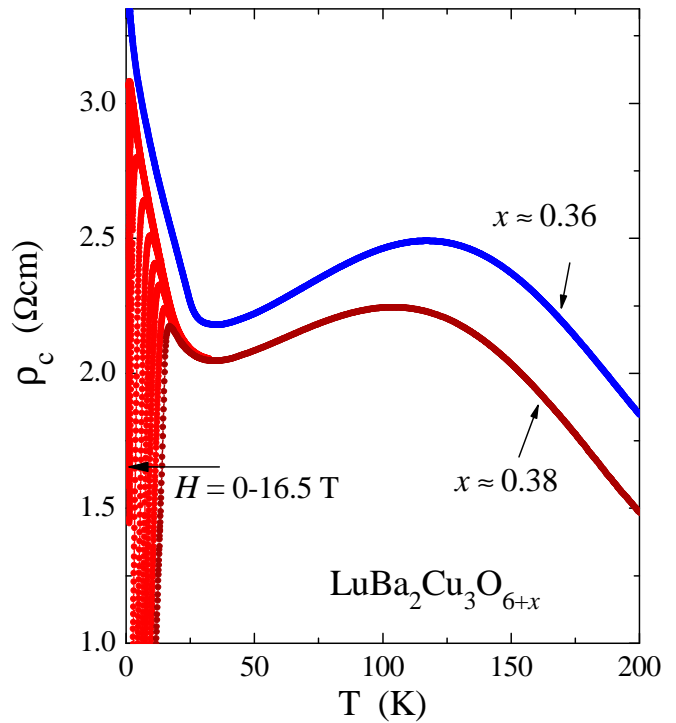}
\caption{Out-of-plane resistivity of one and the same
LuBa$_2$Cu$_3$O$_{6+x}$ single crystal for two oxygen
concentrations near the AF-SC transformation. For the
superconducting composition $x=0.38$, the data were taken at
several magnetic fields--from zero up to 16.5 T--applied along the
$c$ axis. The sharp upturn in the resistivity associated with the
N\'{e}el transition shows up as the superconductivity is
suppressed with the magnetic field.}
\label{fig3}
\end{figure}

According to the established phase diagram (Fig.~2), an increase of the
hole density in CuO$_2$ planes by $\sim \,$1\% per Cu (from $\approx
\,$5\% to $\approx \,$6\%, assuming the onset of superconductivity at
$\approx \,$5\% doping) turns an AF $R$Ba$_2$Cu$_3$O$_{6+x}$ crystal
without any sign of superconductivity into a bulk superconductor with
$T_c \,$$\sim \,$$15 \,$K. What happens with the AF order upon entering
the SC region, does it vanish abruptly? Zero-field $\rho_c(T)$ curves
measured on the same LuBa$_2$Cu$_3$O$_{6+x}$ crystal for two hole doping
levels (that differ by $\approx \,$0.5-0.6\% per Cu) indeed demonstrate
a switch from an AF state with $\rho_c(T)$ sharply growing below $T_N$
to a SC state (Fig.~3). However, when the superconductivity is
suppressed with the magnetic field ${\bf H}$$\,\parallel\,$${\bf c}$,
the steep increase in $\rho_c$ associated with the AF ordering is
recovered back (Fig.~3). Moreover, the recovered N\'{e}el temperature is
merely several Kelvin lower than for the non-SC composition (upper curve
in Fig.~3) and the resistivity increase is not reduced appreciably
either. As long as the SC in LuBa$_2$Cu$_3$O$_{6+x}$ and
YBa$_2$Cu$_3$O$_{6+x}$ crystals is weak enough to be killed by the
16.5-T field, the unveiled $\rho_c(T)$ curves keep demonstrating the
anomalous growth below 15-20$\,$K associated with the N\'{e}el
transition. This behavior indicates that, at least when
superconductivity is suppressed with magnetic fields, the AF order
extends to considerably higher doping levels than the SC onset.
Consequently, at zero magnetic field the AF and SC orders either coexist
with each other in a certain range of doping, or the AF order is
frustrated in the SC state but revives as the superconductivity is
destroyed with the magnetic field. A switching between the AF and SC
orders was indeed suggested based on early $\mu$SR studies
\cite{diagram2}, yet no further proofs were collected. The close
location and even overlapping of the AF and SC orders on the phase
diagram raise another question of whether the AF and SC orders reside in
nanoscopically separated phases in CuO$_2$ planes or coexist on the
unit-cell scale, which calls for local microscopic tools to be
clarified.

\subsection{Acknowledgments}

We thank V. F. Gantmakher for fruitful discussions and acknowledge
support by RFBR (grants 05-02-16973 and 06-02-17098) and the integration
project SB RAS No.81.

\end{document}